\documentclass[showpacs,preprintnumbers,aps, prd]{revtex4}
\usepackage{graphicx}
\usepackage{grffile}
\usepackage{mathrsfs, mathtools}
\usepackage{caption}
\usepackage{float}
\usepackage{xcolor}
\usepackage{changepage}
\usepackage{dcolumn}
\usepackage{bm}
\usepackage{graphicx,subfigure,epsfig}
\usepackage{amsmath,amsfonts,amssymb}
\usepackage{multirow}

\newcommand{\m}{\mathcal{M}}
\def\E{\mathcal{E}}
\def\L{\mathcal{L}_z}
\def\a{\alpha}

\def\f{\frac}
\def\R{\rho^2}
\def\e{\eta}
\def\z{\zeta}
\def\th{\theta}
\def\c{\cite}
\def\r{\ref}

\def\s{Schwarzschild }
\def\k{Kerr }
\def\d{\delta}
\def\D{\Delta}
\newcommand\bt{\bibitem}
\newcommand\be{\begin{equation}}
\newcommand\ee{\end{equation}}
\newcommand\ba{\begin{eqnarray}}
\newcommand\ea{\end{eqnarray}}
\newcommand\nn{\nonumber}
\newcommand\lt{\left}
\newcommand\rt{\right}

\newcommand\tx{\text}
\newcommand\mc{\mathcal}
\begin{document}
\title{Testing linear-quadratic GUP modified Kerr Black hole using EHT results}
\author{Sohan Kumar Jha}
\email{sohan00slg@gmail.com}
\affiliation{Chandernagore College, Chandernagore, Hooghly, West
Bengal, India}

\date{\today}
\begin{abstract}
\begin{center}
Abstract
\end{center}
The linear-quadratic Generalized uncertainty principle (LQG) is consistent with predictions of a minimum measurable length and a maximum measurable momentum put forth by various theories of quantum gravity. The quantum gravity effect is incorporated into a black hole (BH) by modifying its ADM mass. In this article, we explore the impact of GUP on the optical properties of an LQG modified \k BH (LQKBH). We analyze the horizon structure of the BH, which reveals a critical spin value of $7M/8$. BHs with spin $(a)$ less than the critical value are possible for any real GUP parameter $\a$ value. However, as the spin increases beyond the critical value, a forbidden region in $\a$ values pops up that disallows the existence of BHs. This forbidden region widens as we increase the spin. We then examine the impact of $\a$ on the shape and size of the BH shadow for inclination angles $17^o$ and $90^o$, providing a deeper insight into the unified effect of spin and GUP on the shadow. The size of the shadow has a minimum at $\a=1.0M$, whereas, for the exact value of $\a$, the deviation of the shadow from circularity becomes maximum when the spin is less than the critical value. No extrema is observed for $a\,>\, 7M/8$. The shadow's size and deviation are adversely affected by a decrease in the inclination angle. Finally, we confront theoretical predictions with observational results for supermassive BHs $M87^*$ and $SgrA^*$ provided by the EHT collaboration to extract bounds on the spin $a$ and GUP parameter $\a$. We explore bounds on the angular diameter $\th_d$, axial ratio $D_x$, and the deviation from \s radius $\d$ for constructing constraints on $a$ and $\a$. Our work makes LQKBHs plausible candidates for astrophysical BHs.
\end{abstract}
\maketitle
\section{Introduction}
The Heisenberg uncertainty principle (HUP) in quantum mechanics provides a well-known relation between uncertainties in position and momentum, putting no bar on their values. One can make the uncertainty in one as small as zero at the expanse in the uncertainty of the other. However, this needs to be altered significantly near the Planck scale, where we need to amalgamate quantum mechanics and general relativity (GR) into what is known as quantum gravity. A complete theory of quantum gravity is still an open problem. There are various theories of quantum gravity, such as string theory \c{string}, the noncommutative geometry \c{nc}, and loop quantum gravity \c{loop}. All these theories have a common feature: the existence of a minimum measurable length. As one cannot go beyond the minimum length of a string, the existence of a minimum length can readily be understood from the string theory. The loop quantum gravity also entails a minimum length originating from the polymer quantization, where the polymer length plays the role of a minimum length. The presence of minimum length in the loop quantum gravity transforms the big bang into a big bounce \c{loop}. The possibility of a minimum measurable length arises even from the perspective of black hole (BH) physics, where the energy required to probe a spacetime region below the scale of Planck length would produce a mini BH that would make it impossible to investigate the region. It is widely accepted that any theory of quantum gravity must be equipped with a minimum length \c{mm, park}. Theories such as doubly special relativity (DSR), on the other hand, propose a maximum measurable momentum \c{dsr}. \\
To incorporate the existence of a minimum length, a maximum momentum, or both, we need to modify HUP to a Generalized uncertainty principle (GUP). Here, we have an additional contribution to uncertainties arising from gravitational effects. There are various GUPs in the literature that are consistent with predictions of either string theory, DSR, or both of them. For example, the quadratic GUP (QG) \c{qg1} ensures the existence of a minimum length, thus consistent with string theory. The linear GUP (LG) \c{lq}, on the other hand, ensures the presence of a maximum momentum and is thus consistent with DSR. The presence of both a minimum measurable length and a maximum measurable momentum apparently leads to a GUP, namely linear-quadratic GUP (LQG), that unifies linear and quadratic GUPs \c{lqg1, lqg2, lqg3, lqg4}. An important consequence of GUP is the presence of BH remnants where the backreaction effects stop the evaporation process near the Planck scale \c{dm, dm2, dm3}. These remnants, if found in significant number, can be a plausible candidate of dark matter. The effect of GUP is incorporated into BH by modifying its ADM mass accordingly. We have studied the optical properties of LQG modified \k BH (LQKBH) in this article. Please see [\citenum{31}-\citenum{53}] and references therein for studies of different properties of GUP-modified BHs.\\
The remarkable feat achieved by the EHT collaboration in reconstructing images of supermassive BHs (SMBHs) $M87^*$ and $SgrA^*$ [\citenum{m87}-\citenum{sgra1}] and the detection of gravitational waves by LIGO and VIRGO \c{ligo} have proved an astounding prediction of GR: the existence of BHs. These remarkable discoveries have generated significant interest in the study of BH. BHs are endowed with strong gravitational pull at event horizons and nearby photon regions. This results in a dark region, known as a shadow, in the celestial sky of the observer \c{bardeen, shadow}. It is outlined by a bright ring formed by gravitationally lensed photons. A BH shadow depends on the intrinsic parameters such as spin, mass, and hair of BH. As such, they encode the fundamental characteristics of the background spacetime. This has made the BH shadow the central object of intense research [\citenum{8}-\citenum{jha}]. Bounds on various observables related to the shadow size and shape of SMBHs $M87^*$ and $SgrA^*$ extracted from the EHT observations have presented an excellent avenue to confront theoretical predictions with experimental results. A significant number of studies has been devoted to testing different classes of BHs arising from various gravity theories with the help of the EHT observations [\citenum{74}-\citenum{114}], thereby enriching our understanding and providing deeper insights into these theories. Our central aim in this work is to constrain spin and GUP parameters by utilizing bounds on various observables related to $M87^*$ and $SgrA^*$ shadows given by the EHT collaboration.\\
We organize our work in this article as follows. In Sec. II, we study the horizon structure of LQKBHs and obtain conditions to be put on the values of spin and GUP parameters for the existence of a BH solution. By utilizing the Hamilton-Jacobi equation, we then derive differential equations for null geodesics and explore shadow properties by varying spin and GUP parameters in Sec. III. Subsequently, in Sec. IV, we compare theoretical predictions with EHT observational data for both SMBHs. The final section concludes our article with a concise overview of our findings and a proposed future endeavor.

\section{LQG modified \k BH}
The following QG ensures minimum measurable length predicted by various quantum gravity theories such as string theory and black hole physics \c{qg1}:
\be
\D x\D p\geq \f{\hbar}{2}\lt[1+\a^2 \D p^2 \rt],
\ee
where $\a =\a_0 /M_{pl}c=a_0 \ell_{pl}/\hbar$, $M_{pl}$ and $\ell_{pl}$ being Planck mass and length, respectively. The above GUP implies the minimum measurable length to be $\D x_{min} \approx \a_{0} \ell_{pl}$. DSR theories, on the other hand, propose a maximum measurable momentum \c{dsr}. Following is the LG that is consistent with DSR theories providing a maximum measurable momentum \c{lq}:
\be
\D x\D p\geq \f{\hbar}{2}\lt[1-\a \D p \rt],
\ee
where the maximum measurable momentum is $\D p_{max} \approx \f{M_{pl}c}{\a_{0}}$. The LQG combines the two GUPs to produce the effect of a minimum measurable length and a maximum measurable momentum. We take the following form of LQG for our purpose in this manuscript:
\be
\D x\D p\geq \f{\hbar}{2}\lt[1-\a \D p +\a^2 \D p^2\rt],
\label{lqg}
\ee
which ensures minimum measurable length and a maximum measurable momentum given by
\be
\D x \geq \D x_{min} \approx \a_{0} \ell_{pl}\quad \tx{and} \quad \D p \leq \D p_{max} \approx \f{M_{pl}c}{\a_{0}}.
\ee
Owing to the LQG (\r{lqg}), the mass of the BH gets modified according to \c{lqg1, lqg2, lqg3, lqg4}
\be
\m = M_{LQG}=M\lt(1-\f{\a}{4M}+\f{\a^2}{8M^2}\rt),
\ee
where $\m$ is the ADM mass and $M$ is the bare mass of BH. As such, the metric for the LQG modified \k BH is given by
\ba
ds^{2}&=&-\left[\frac{\Delta-a^{2}\sin^{2}\theta}{\R}\right]dt^{2}+\frac{\R}{\Delta}dr^{2}+\R d\theta^{2}-2a\sin^2{\th}\left[1-\frac{\Delta-a^2\sin^2{\th}}{\R}\right]dtd\phi \nonumber\\
&&+\frac{\sin^2{\theta} }{\R}\left[\left(r^{2}+a^2\right)^2-\Delta a^2\sin^2{\theta} \right]d\phi^{2}, \\
\text{with}&&\nonumber\\
\Delta&=&a^{2}+r^2-2\m r,\quad \R=r^2+a^{2}\cos^{2}\theta.\label{metric}
\ea
In the limit $\a \rightarrow 0$, we recover \k metric, whereas, in the limit $\a \rightarrow 0$ and $a \rightarrow 0$, we get back the metric for the \s BH. Metric (\r{metric}), similar to the \k BH, has two singularities. One is the ring singularity at $\R=0$; the other is the coordinate singularity at $\D=$ and $\R \neq 0$. The solution of the equation $\D=g^{rr}=0$ provides radii of the event horizon $(r_{eh})$ and Cauchy horizon $(r_{ch})$. They are given by
\ba\nn
r_{eh}&=&\m+\sqrt{\m^2-a^2}=M\lt(1-\f{\a}{4M}+\f{\a^2}{8M^2}\rt)+\sqrt{M^2 \lt(1-\f{\a}{4M}+\f{\a^2}{8M^2}\rt)^2-a^2},\\
r_{ch}&=&\m-\sqrt{\m^2-a^2}=M\lt(1-\f{\a}{4M}+\f{\a^2}{8M^2}\rt)+\sqrt{M^2 \lt(1-\f{\a}{4M}+\f{\a^2}{8M^2}\rt)^2-a^2}.\\
\ea
There exists a BH solution when the expression within the square bracket in the above expression is greater than or equal to zero, providing a relation between the GUP parameter $\a$ and spin $a$ for the existence of a BH solution. The required condition is
\be
M\lt(1-\f{\a}{4M}+\f{\a^2}{8M^2}\rt) \geq a. \label{condition}
\ee
The above condition puts constrain on the possible values of $a$ and $\a$. When the equality in the above condition is satisfied, the two horizons coincide, and we get extremal BH. When the spin parameter $a < \f{7M}{8}$, the inequality in (\r{condition}) is satisfied for all real values of $\a$, but the equality is never satisfied. It implies that for no real value of $\a$, we will have extremal BH when $a < \f{7M}{8}$. For the critical value of the spin parameter, i.e., $a=\f{7M}{8}$, the equality is satisfied for $\a=1.0M$ and we have an extremal BH. Thus, for all real values of $\a$, we have a BH solution even at the critical spin.
\begin{figure}[H]
\centering
\includegraphics[width=0.4\columnwidth]{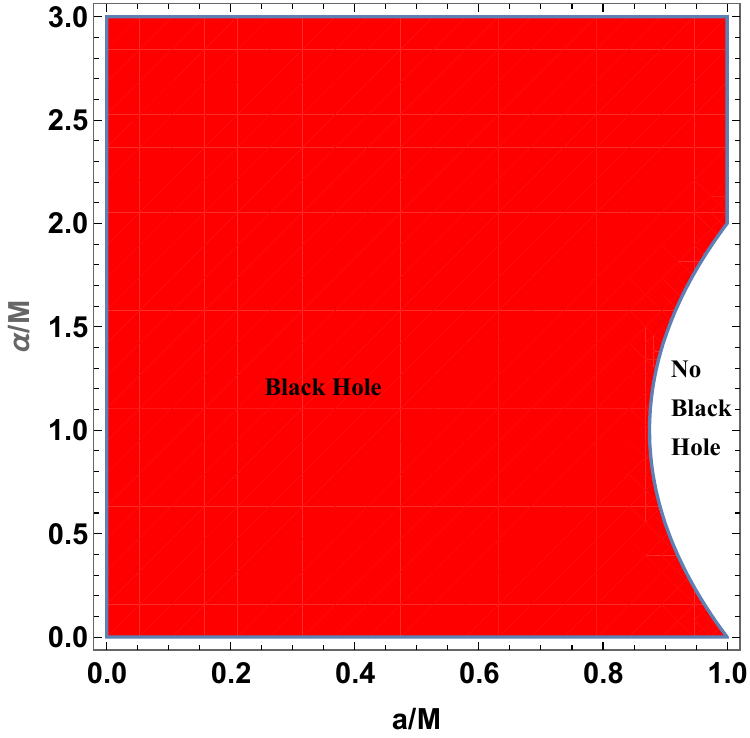}
\caption{Parameter space $(a-\a)$ for the LQKBH where the parameter values in the colored region permit the BH solution, and the white region is where the BH solution is disallowed.}
\label{regionplot}
\end{figure}
When the spin parameter exceeds the critical value of $7M/8$, a forbidden region in terms of GUP parameter values pops up, disallowing the BH solution. For a spin $a> \f{7M}{8}$, forbidden values of $\a$ lie in the open interval $\lt(M-\sqrt{8 a M-7 M^2}, M+\sqrt{8 a M-7 M^2}\rt)$. In this case, BH solutions are permitted for $\a \leq M-\sqrt{8 a M-7 M^2}$ and $\a \geq M+\sqrt{8 a M-7 M^2}$. We get extremal BHs for $a> \f{7M}{8}$ at $\a=M-\sqrt{8 a M-7 M^2}$ and $\a=M+\sqrt{8 a M-7 M^2}$. An increase in the spin parameter enlarges the forbidden interval. For example, if we take $a=0.9M$, the interval of $\a$ with no BH solution becomes $(0.552786M, 1.44721M)$. At $\a=0.552786M$ and $\a=1.44721M$ for $a=0.9M$, the event horizon and the Cauchy horizon coincide at the same value for both the GUP parameters. Values of the GUP parameter $\a$ that generate no BH solution lie in the interval $(0,2)$ for spin $M$. Fig. (\r{del}) illustrates the impact of the spin parameter $a$ and GUP parameter $\a$ on the horizons of the BH. Here, we have taken two spin parameters: one less than the critical value and the other greater than the critical value.
\begin{figure}[H]
\begin{center}
\begin{tabular}{cc}
\includegraphics[width=0.4\columnwidth]{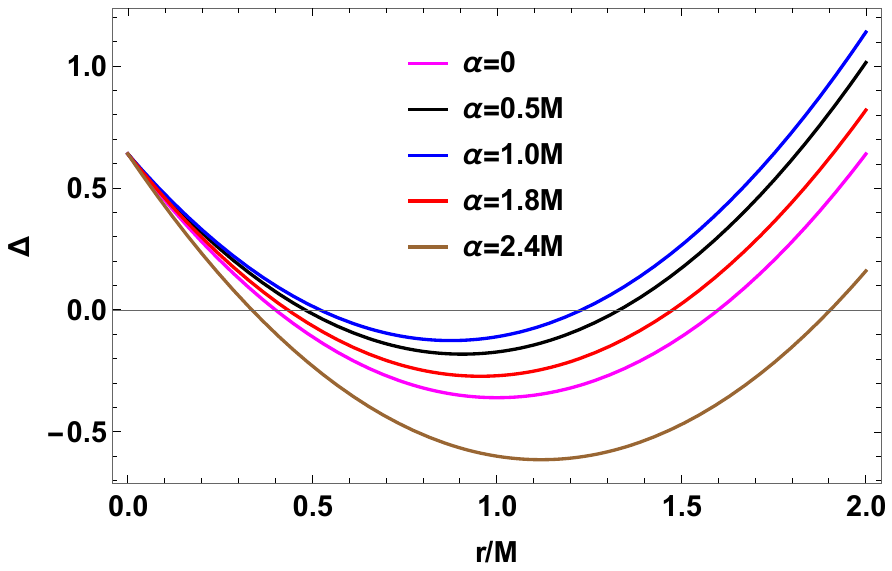}&
\includegraphics[width=0.4\columnwidth]{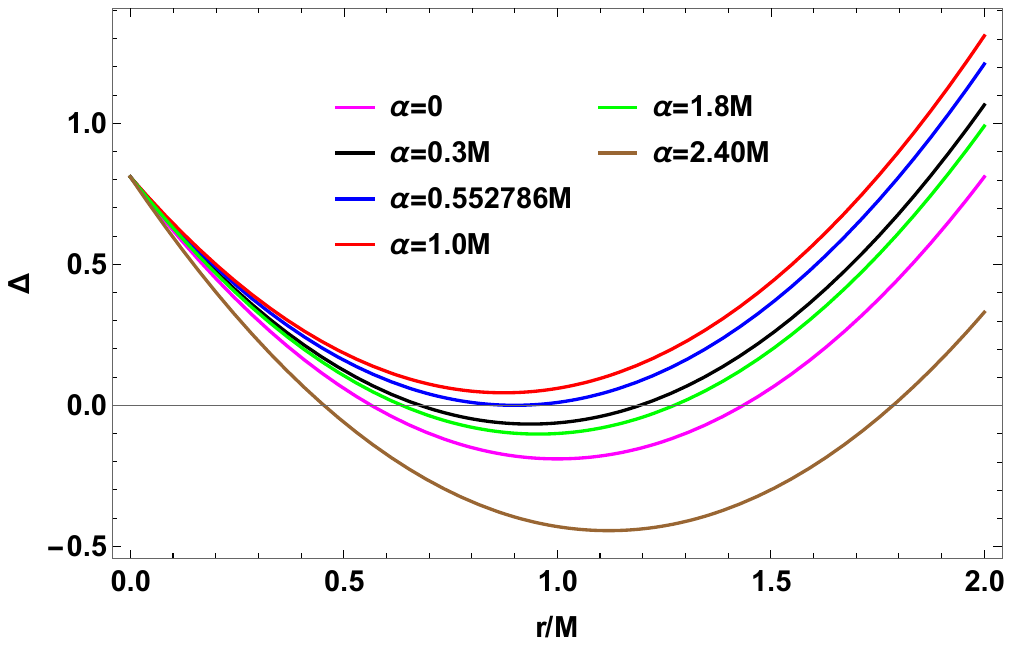}
\end{tabular}
\caption{Variation of $\D$ with $r/M$ for different values of $\a$. The left one is for spin $a=0.8M$ and the right one is for spin $a=0.9M$.}\label{del}
\end{center}
\end{figure}
Two horizons demonstrate critical behavior when the spin is less than $7M/8$. When the GUP parameter increases, with spin fixed at $a=0.8M$, initially, the event horizon decreases, and the Cauchy horizon increases till $\a=1.0M$. When the GUP parameter is increased further, the event horizon increases, and the reverse happens for the Cauchy horizon. However, no such turning point can be observed for spin parameter $a=0.9M$ as we do not have BH for $\a=1.0M$. In this case, the event horizon decreases, and the Cauchy horizon increases with an increase in $\a$ till $\a < 0.552786M$, and the trend gets reversed for $\a > 1.44721M$.
\section{black hole shadow}
The black hole shadow is attributed to null geodesics. It is a two-dimensional dark region on the observer's sky outlined by gravitationally lensed photons. Null geodesics solely depend on the characteristics of the underlying spacetime. As a result, we can extract valuable information related to the background spacetime encoded by black hole shadows. To this end, we need first to obtain differential equations governing null geodesics. Metric (\r{metric}) evinces two killing vectors: one is $\d^{\mu}_{t}$ corresponding to time translational isometry, and another one is $\d^{\mu}_{\phi}$ corresponding to rotational invariance isometry. As such, energy $\E=-g_{t\phi}\dot{\phi}-g_{tt}\dot{t}$ and azimuthal component of angular momentum $\L=g_{\phi\phi}\dot{\phi}+g_{\phi t}\dot{t}$ are constants of motion along the path of photons. Additionally, a third constant, namely Carter constant $\mc{Q}$, arises from the separability of the Hamilton-Jacobi equation first shown by Carter \c{carter, chandra}. Following is the Hamilton-Jacobi equation:
\begin{equation}
\frac{\partial \mathcal{S}}{\partial \lambda}=-\frac{1}{2}g^{\mu\nu}\frac{\partial \mathcal{S}}{\partial x^\mu}\frac{\partial \mathcal{S}}{\partial x^\nu},
\label{hje}
\end{equation}
where $\lambda$ is the affine parameter and $\mathcal{S}$ is the Jacobi action. The action of the form
\begin{equation}
\mathcal{S}=\frac{1}{2}m_{0} ^2 \lambda - \E t +\L \phi + \mathcal{S}_{r}(r)+\mathcal{S}_{\theta}(\theta),
\label{action}
\end{equation}
renders Eq. (\r{hje}) separable into four first-order differential equations given by:
\begin{align}
\R \dot{t}=&\frac{r^2+a^2}{\Delta}\left[\mathcal{E}\left(r^2+a^2\right)-a\mathcal{L}_z\right]\nonumber\\
&-a(a\mathcal{E}\sin^2{\theta}-\mathcal{L}_z),\label{teq}\\
\R \dot{\phi}=&\frac{a}{\Delta}\left[\mathcal{E}\left(r^2+a^2\right)-a\mathcal{L}_z\right]-\left(a\mathcal{E}-\frac{\mathcal{L}_z}{\sin^2{\theta}}\right),\label{phieq}\\
\R \dot{r}=&\sqrt{\Big[\left(r^2+a^2\right)\mathcal{E}-a \mathcal{L}_z \Big]^2-\Delta \Big[{\mathcal{K}}+(a \mathcal{E}- \mathcal{L}_z)^2\Big]}\nonumber\\
&\equiv\sqrt{\mathcal{R}(r)}\ ,\label{req} \\
\R \dot{\theta}=&\sqrt{\mathcal{K}-\left(\frac{{\mathcal{L}_z}^2}{\sin^2\theta}-a^2 \mathcal{E}^2 \right)\cos^2\theta}\equiv\sqrt{\Theta(\theta)}\ .\label{theq}
\end{align}
$m_0$ in Eq. (\r{action}) is the test particle's rest mass, which is zero for photons. Overdot in the above equations represents differentiation with respect to the affine parameter $\lambda$, and $\mathcal{K}=\mathcal{Q}-(a\E-\L)^2$ is the separability constant. We recover null geodesics equations for \k BH in the limit $\a \rightarrow 0$. BH shadows are attributed to the unstable circular photon orbits whose radii $r_p$ are obtained by imposing the following conditions on the radial potential $\mc{R}(r)$:
\be
\mc{R}(r_p)=\f{d\mathcal{R}}{d\text{r}}|_{r_p}=0 \quad \text{and} \quad \f{d\mathcal{R}^2}{d\text{r}^2}|_{r_p} > 0.
\ee
Solving the above equations yields the following impact parameters for null geodesics in unstable circular orbits:
\begin{align}
\z=&\frac{\left(a^2+r^2\right) \Delta '(r)-4 r \Delta (r)}{a \Delta '(r)},\nonumber\\
\e=&\frac{r^2 \left(8 \Delta (r) \left(2 a^2+r \Delta '(r)\right)-r^2 \Delta '(r)^2-16 \Delta (r)^2\right)}{a^2 \Delta '(r)^2}\label{ip},
\end{align}
Unstable circular orbits lies between the prograde orbit of radius $(r_{-})$ ($\z > 0$) and the retrograde orbit of radius $r_{+}$ ($\z < 0$) where $r^{\pm}$ are solutions of equation $\e=0$. These orbits form a photon shell whose projection onto the celestial sky of an asymptotic observer provides the BH shadow. A BH shadow is a dark region outlined by the following celestial coordinates \c{bardeen}:
\begin{equation}
\{X,Y\}=\{-\z \csc\theta_o,\, \pm\sqrt{\eta+a^2\cos^2\theta_o-\z^2\cot^2\theta_o}\}\,\label{celestial}
\end{equation}
where $\theta_{o}$ is the inclination angle relative to the spin axis.
\begin{figure}[H]
\begin{center}
\begin{tabular}{cc}
\includegraphics[width=0.4\columnwidth]{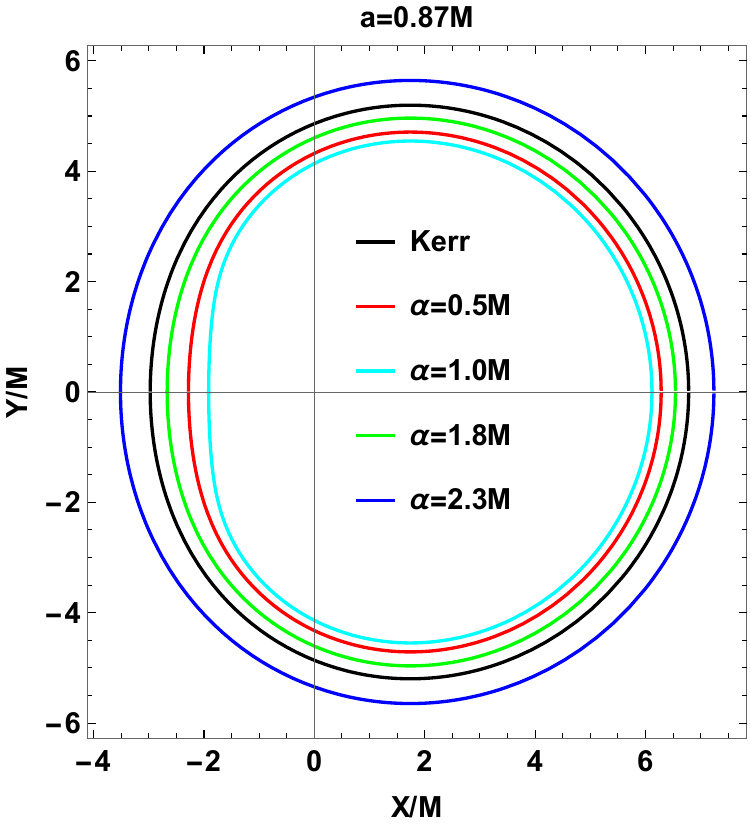}&
\includegraphics[width=0.4\columnwidth]{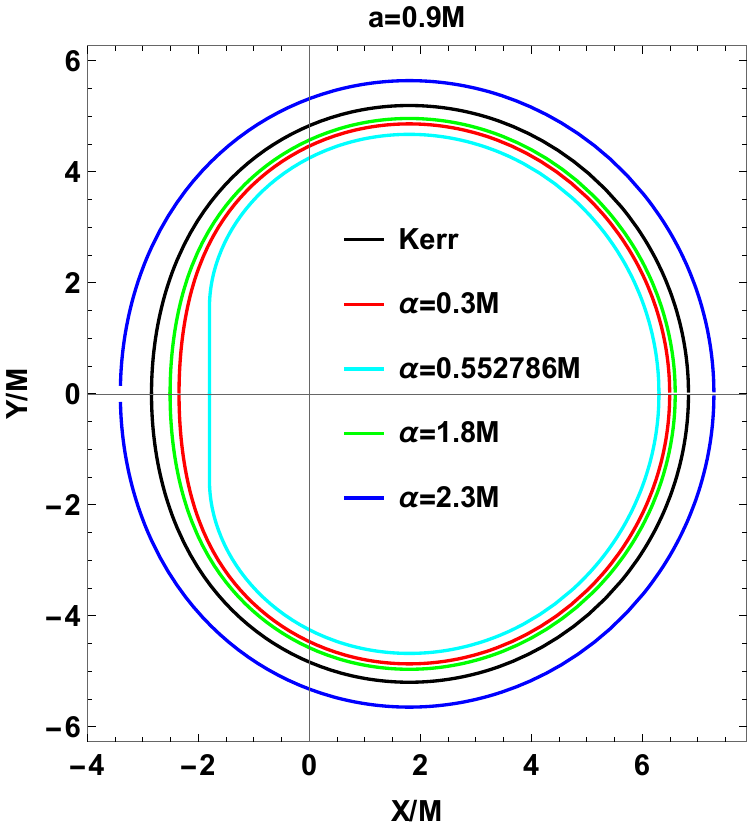}
\end{tabular}
\caption{Shadows cast by LQKBH when the inclination angle is $\th_o=90^o$. The left one is for $a=0.87M$ and the right one is $a=0.90M$. }\label{shadow90}
\end{center}
\end{figure}
\begin{figure}[H]
\begin{center}
\begin{tabular}{cc}
\includegraphics[width=0.4\columnwidth]{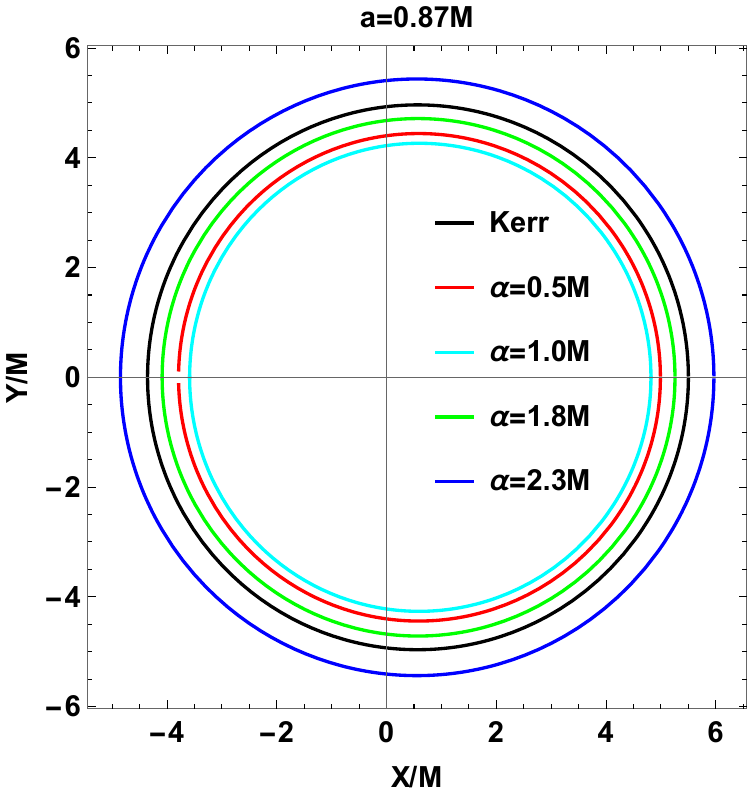}&
\includegraphics[width=0.4\columnwidth]{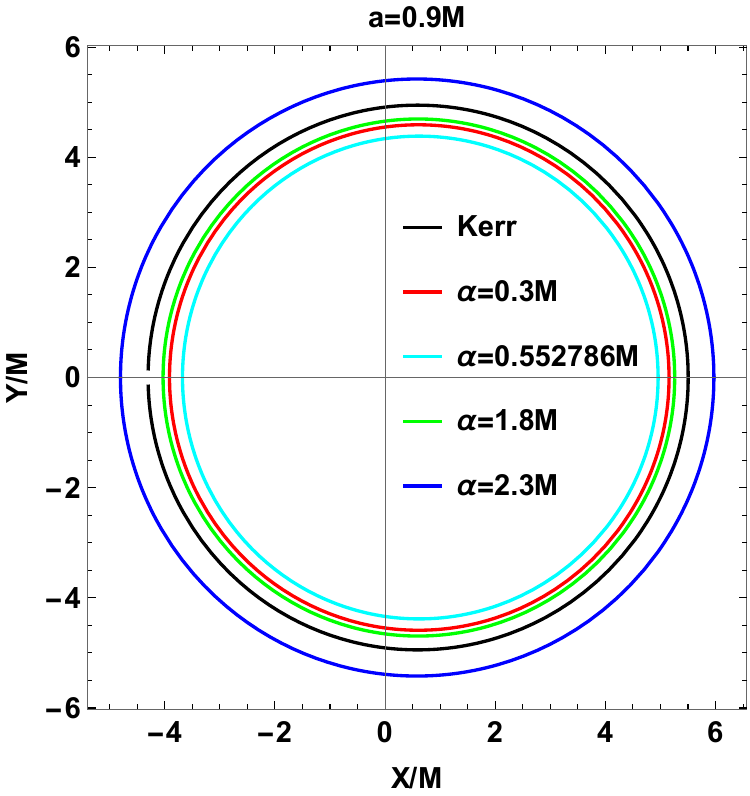}
\end{tabular}
\caption{Shadows cast by LQKBH when the inclination angle is $\th_o=17^o$. The left one is for $a=0.87M$ and the right one is $a=0.90M$. }\label{shadow17}
\end{center}
\end{figure}
Figs. (\r{shadow90}) and (\r{shadow17}) exhibit the influence of the GUP parameter $\a$ on the size and shape of the shadow cast by an LQKBH. Similar to the case of event horizon, the size of the shadow initially decreases with $\a$ reaching a minimum at $\a=1.0M$ and then starts increasing with $\a$ when $a<7M/8$. On the contrary, the distortion of the shadow shape increases with $\a$ till $\a=1.0M$ and then starts decreasing. However, no such minimum in the shadow size is observed for $\a>7M/8$. Here, the shadow size decreases with $\a$ when $\a <  M-\sqrt{8 a M-7 M^2}$ and increases with $\a$ when $\a >  M+\sqrt{8 a M-7 M^2}$. Reverse is the situation with regard to the deviation of the shadow from the circular shape. A comparison of Figs. (\r{shadow90}) and (\r{shadow17}) reveals that a decrease in the inclination angle adversely impacts both the shadow's size and distortion.
\section{Testing viability of LQKBHs using EHT results}
The extraction of constraints on the parameters of BHs is one of the most important problems in astrophysics. With the images of SMBHs $M87^{*}$ \c{m87, m871} and $SgrA^{*}$ \c{sgra, sgra1} observed by the EHT collaboration \c{m87, m871, sgra, sgra1} open up an unparalleled opportunity to probe various theories of quantum gravity. Owing to significant uncertainties in the results produced by EHT, even though the shadow cast a \k BH agrees well with observations, we cannot rule out theories of quantum gravity. The EHT collaboration has provided bounds on various observables that quantify shadow size and shape. We will be employing bounds on the angular diameter $\th_d$, deviation from \s radius $\d$, and the axial ratio $D_x$ to put constrain on the possible values of spin and GUP parameters by modeling the two SMBHs as LQKBH.
\subsection{Obtaining bounds from the angular diameter and axial ratio}
Shadows of SMBHs $M87^{*}$ and $SgrA^{*}$ observed by the EHT collaboration are outlined by asymmetric bright rings. Their angular diameters, along with their inferred distances from the Earth $D$ and masses, are \c{m87, m871, sgra, sgra1}:
\ba
&&\text{$M87^*$: } M=6.5\times10^9 M_\odot, \quad D=16.8Mpc, \quad \text{and} \quad \th_{d}=42\pm3 \mu as, \\
&&\text{$SgrA^*$:} M= 4.0 \times 10^6 M_\odot, \quad D=8 kpc, \quad \text{and} \quad \th_{d}=47.8\pm7 \mu as.
\ea
We can calculate theoretical values of the angular diameter using the formula \c{area}
\ba
\th_d=2\f{R_a}{D}, \quad \text{where}\quad
R_a=\sqrt{\f{A}{\pi}}, \quad \quad A=2\int_{r_{-}}^{r_+}\left( Y \frac{dX}{dr}\right)dr.\label{def}
\ea
Here $A$ is the shadow area and $R_a$ is the shadow areal radius. The inclination angle inferred by EHT for $M87^*$ is $17^o$. We will carry out our analysis for inclination angles $17^o$ and $90^o$.
\begin{figure}[H]
\begin{center}
\begin{tabular}{cccc}
\includegraphics[width=0.4\columnwidth]{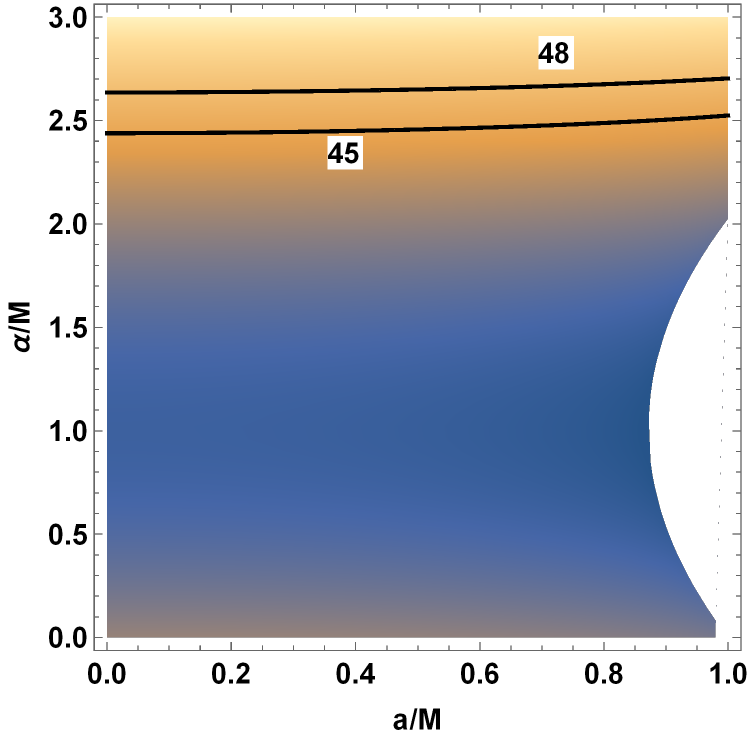}&
\raisebox{.055\height}{\includegraphics[width=0.04\columnwidth]{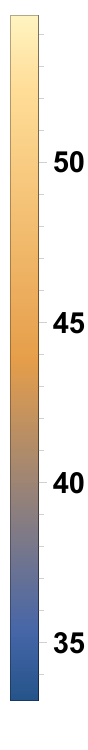}}&
\includegraphics[width=0.4\columnwidth]{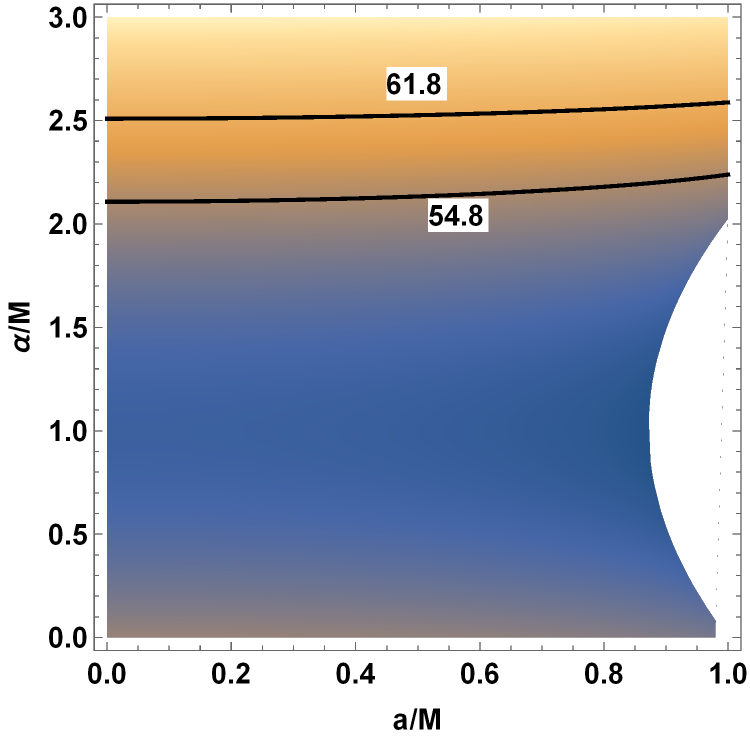}&
\raisebox{.055\height}{\includegraphics[width=0.04\columnwidth]{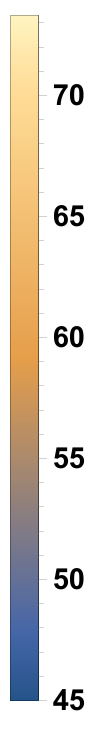}}\\
\includegraphics[width=0.4\columnwidth]{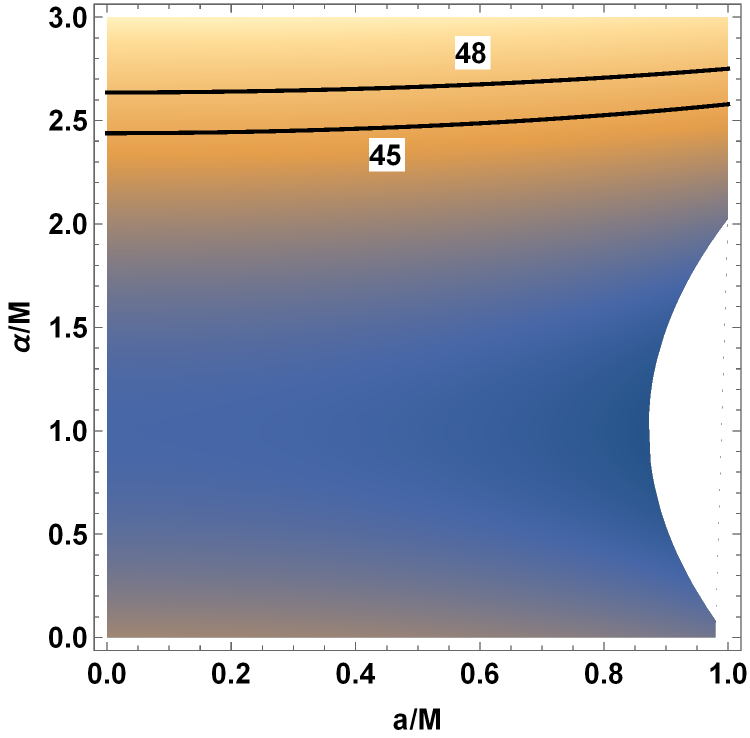}&
\raisebox{.055\height}{\includegraphics[width=0.04\columnwidth]{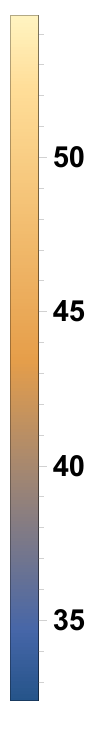}}&
\includegraphics[width=0.4\columnwidth]{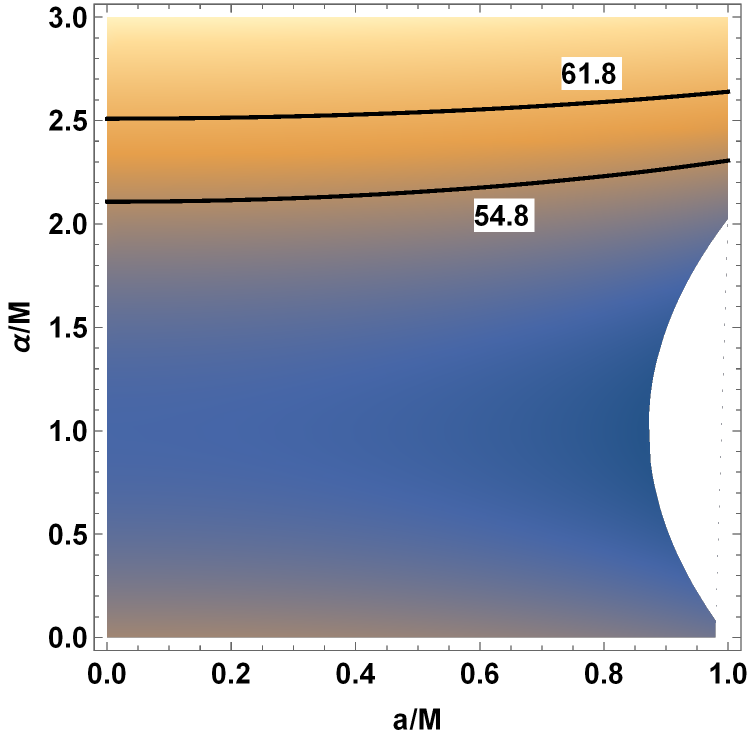}&
\raisebox{.055\height}{\includegraphics[width=0.04\columnwidth]{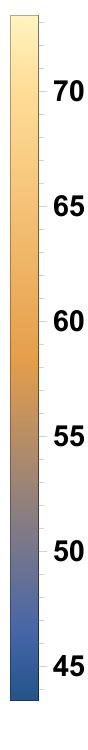}}
\end{tabular}
\caption{Variation of angular diameter $\th_d$ with spin $a$ and GUP parameter $\a$. Here, the values along bold lines and sidebars are in $\mu as$. The upper panel is for $\th_0=90^o$, and the lower panel is for $\th_o=17^o$. Left plots are for $M87^*$ and right plots are for $SgrA^*$.}\label{thetapara}
\end{center}
\end{figure}
Fig. (\r{thetapara}) shows that for a sizeable part of the parameter space $(a, \a)$, our model is concordant with the EHT observations related to the angular diameter of shadow. Bouds on the spin is found to be $[0, M]$. However, the inclusion of possible values of spin greater than the critical value $7M/8$ creates a range of values of GUP parameter $\a$ that does not generate a BH solution. This range is maximum when $a=M$ where the BH solution is allowed for $\a=0$ and $\a \geq 2$. We here report upper bounds on $\a$ $(\a_{u})$, which indicate that apart from the possibility of $\a=0$ (i.e., \k BH), GUP parameter values that produce theoretical values consistent with the EHT observations lie in the set $[2, \a_u]$. Obtained upper bounds are tabulated below:
\begin{center}
\begin{tabular}{|l|c|c|c|c|}
\hline
BH & Inclination angle  & 1$\sigma$ upper bound & 2$\sigma$ upper bound\\
\hline
\multirow{2}{*}{$M87^*$} &{$17^o$} & {$ 2.43918M$} & {$2.6356M$}\\[3mm]
&$90^o$& $2.439208M$ & $2.63584M$\\
\hline
\multirow{2}{*}{$SgrA^*$} &{$17^o$} & {$ 2.01765M$} & {$2.50872M$}\\[3mm]
&$90^o$& $2.01774M$ & $2.50932M$\\
\hline
\end{tabular}
\captionof{table}{Upper Bounds on $\a$ from bounds on the angular diameter.} \label{bounds_angle}
\end{center}
Table (\r{bounds_angle}) indicates that the upper bounds obtained from $SgrA^*$ are more stringent than those obtained from $M87^*$. Apart from the size of the shadow that is quantified by the angular diameter, we can quantify the deviation of the shadow from circularity using the axial ratio, which is the ratio of major to minor diameters of the shadow \c{m87}. It is defined by \c{axial}:
\be
D_{x}=\f{\D Y}{\D X},
\ee
where $\D Y$ and $\D X$ are major and minor diameters of shadow. According to the EHT observation, the axial ratio for $M87^*$ lies in the range $1< D_x\lesssim 1.33$. We exhibit variation of the axial ratio with the spin and the GUP parameter in Fig. (\r{axialpara}). It shows that the EHT bounds on the axial ratio are satisfied by the LQKBH for the entire parameter space $(a,\, \a)$.
\begin{figure}[H]
\begin{center}
\begin{tabular}{cccc}
\includegraphics[width=0.35\columnwidth]{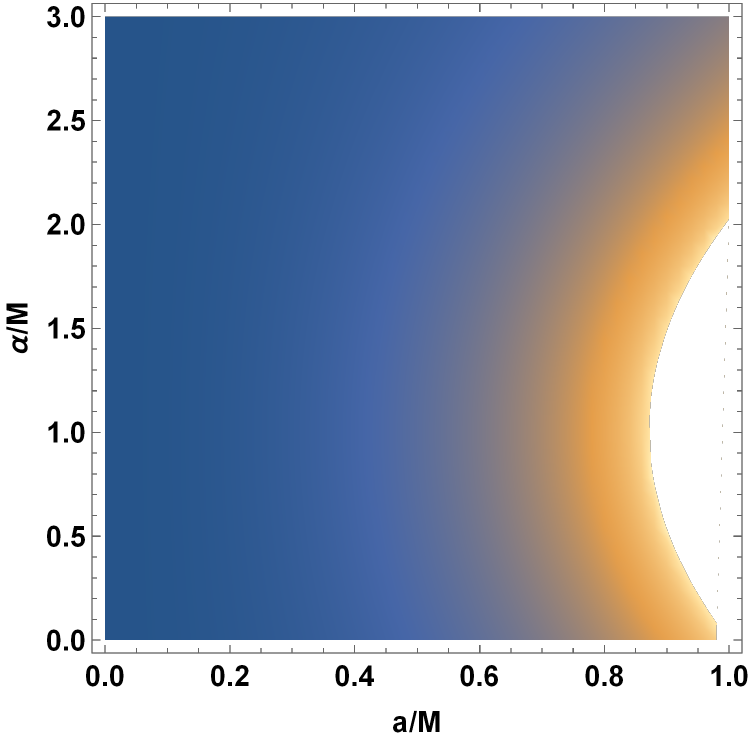}&
\raisebox{.055\height}{\includegraphics[width=0.05\columnwidth]{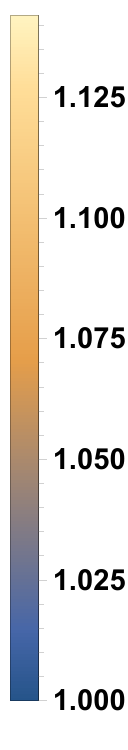}}&
\includegraphics[width=0.35\columnwidth]{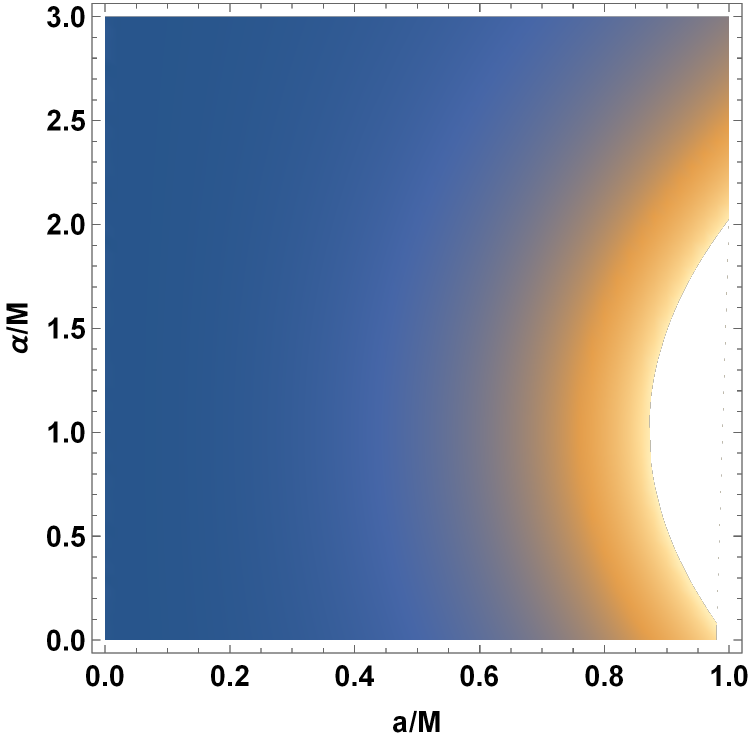}&
\raisebox{.055\height}{\includegraphics[width=0.056\columnwidth]{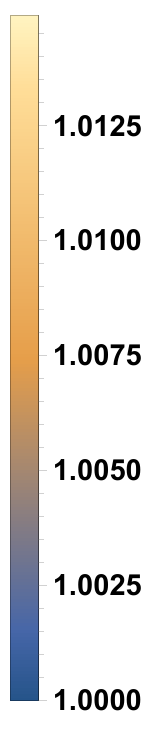}}
\end{tabular}
\caption{Variation of the axial ratio $D_x$ with spin $a$ and GUP parameter $\a$. The left one is for $\th_0=90^o$, and the right one is for $\th_o=17^o$. }\label{axialpara}
\end{center}
\end{figure}
\subsection{Obtaining bounds by utilizing the deviation from \s radius data}
EHT bounds on the deviation from the \s radius, $\d$, equip us with yet another potent tool to test the viability of our model as an SMBH. The deviation is defined as \c{sgra, sgra1}
\be
\d = \f{R_a}{3\sqrt{3}M}-1,
\ee
where $R_a$ is the shadow areal radius given in Eq. (\r{def}). Bounds on $\d$ for $M87^*$ \c{m871} and $SgrA^*$ \c{sgra, sgra1} are tabulated below:
\begin{center}
\begin{tabular}{|l|c|c|c|r|}
\hline
BH & Observatory & $\delta$ \\
\hline
$M87^*$ & EHT & $-0.01^{+0.17}_{-0.17}$ \\
\hline
\multirow{2}{*}{$Sgr A^*$}&{VLTI} & $-0.08^{+0.09}_{-0.09}$ \\[3mm]
& Keck & $-0.04^{+0.09}_{-0.10}$ \\[1mm]
\hline
\end{tabular}
\captionof{table}{Bounds on $\delta$  for $M87^*$ and $SgrA^*$.} \label{delbounds}
\end{center}
The deviation parameter lies in the $[-0.075,\, 0]$ range for a \k BH, agreeing well with the observed bounds. However, significant uncertainties in the observed values leave room for modified theories of gravity. We model SMBHs $M87^*$ and $SgrA^*$ as LQKBH to constrain parameters $a$ and $\a$. Similar to our analysis with regard to the angular diameter, here too, we will report upper bounds on $\a$. The bounds on the spin parameter are $[0,\, M]$. Variations of the deviation parameter against spin and GUP parameter are shown in Figs. (\r{delpara},\r{delpara1}, \r{delpara2}) for inclination angles $17^o$ and $90^o$. Shadow of a LQKBH is smaller as well as larger than a \s BH, as evident from negative and positive values of $\d$ in Figs. (\r{delpara}), (\r{delpara1}), and (\r{delpara2}).
\begin{figure}[H]
\begin{center}
\begin{tabular}{cccc}
\includegraphics[width=0.35\columnwidth]{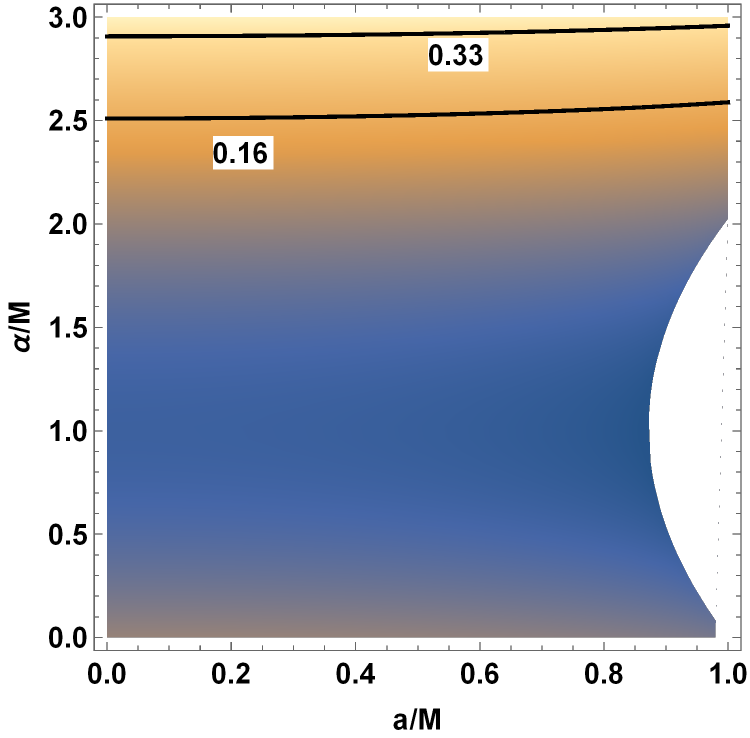}&
\raisebox{.055\height}{\includegraphics[width=0.04\columnwidth]{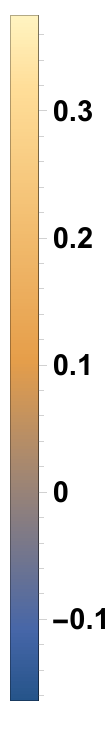}}&
\includegraphics[width=0.35\columnwidth]{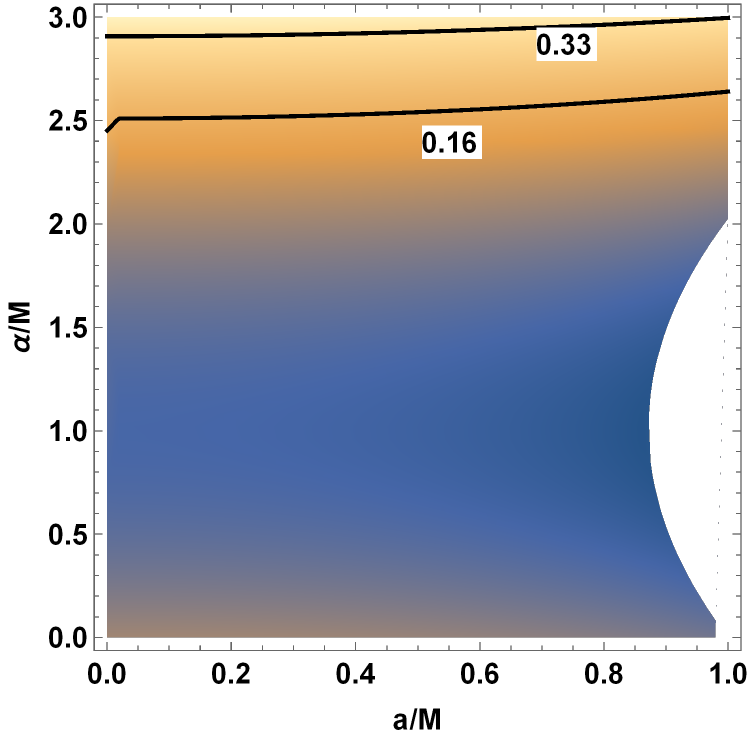}&
\raisebox{.055\height}{\includegraphics[width=0.04\columnwidth]{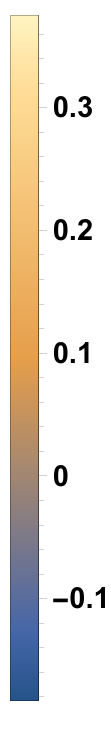}}
\end{tabular}
\caption{Variation of deviation parameter $\d$ with spin $a$ and GUP parameter $\a$. The left one is for $\th_0=90^o$, and the right one is for $\th_o=17^o$. Here, we have taken $M87^*$ and the bold lines provide bounds given by EHT.}\label{delpara}
\end{center}
\end{figure}
\begin{figure}[H]
\begin{center}
\begin{tabular}{cccc}
\includegraphics[width=0.35\columnwidth]{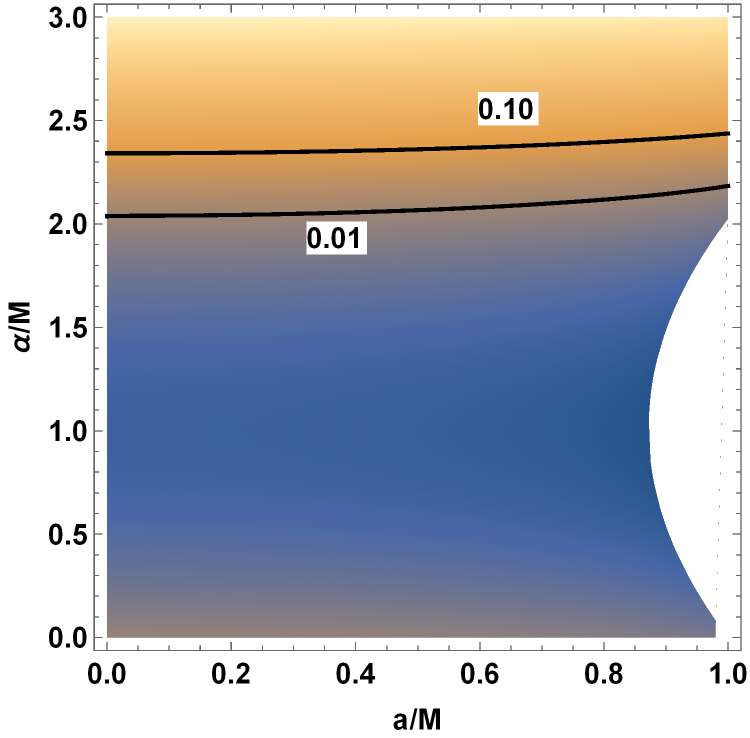}&
\raisebox{.055\height}{\includegraphics[width=0.04\columnwidth]{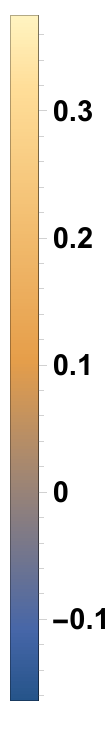}}&
\includegraphics[width=0.35\columnwidth]{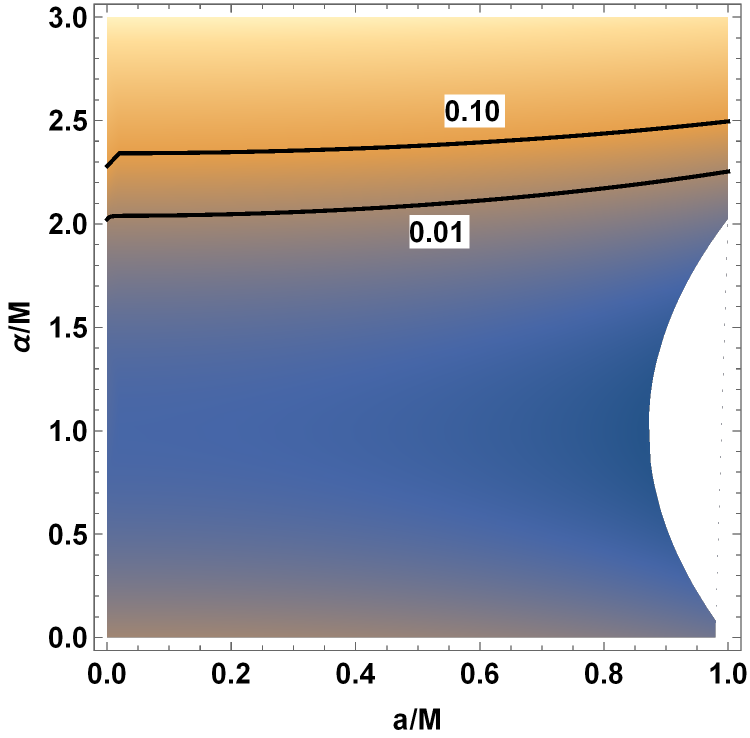}&
\raisebox{.055\height}{\includegraphics[width=0.04\columnwidth]{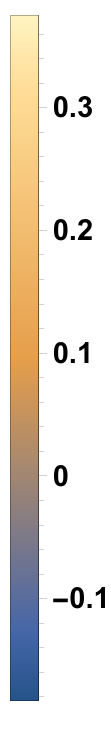}}
\end{tabular}
\caption{Variation of deviation parameter $\d$ with spin $a$ and GUP parameter $\a$. The left one is for $\th_0=90^o$, and the right one is for $\th_o=17^o$. Here we have taken $SgrA^*$ and the bold lines provide bounds given by VLTI.}\label{delpara1}
\end{center}
\end{figure}
It is evident from these figures that our model, for a finite parameter space $(a,\, \a)$, commensurates with observed results. We retrieve the following upper bounds on the GUP parameter:\\
(i) 2.509971M within $1\sigma$ bounds, 2.90787M within $2\sigma$ bounds at $17^*$ and 2.509975M within $1\sigma$ bounds, 2.90788M within $2\sigma$ bounds at $90^*$ following EHT results for $M87^*$. (ii) 2.18312M within $1\sigma$ bounds, 2.45595M within $2\sigma$ bounds at $17^*$ and 2.18323M within $1\sigma$ bounds, 2.455986M within $2\sigma$ bounds at $90^*$ following Keck results for $SgrA^*$. (iii) 2.03921M within $1\sigma$ bounds, 2.34149M within $2\sigma$ bounds at $17^*$ and 2.03925M within $1\sigma$ bounds, 2.34164M within $2\sigma$ bounds at $90^*$ following VLTI results for $SgrA^*$. Bounds given by VLTI result in the most stringent upper bounds on $\a$.
\begin{figure}[H]
\begin{center}
\begin{tabular}{cccc}
\includegraphics[width=0.35\columnwidth]{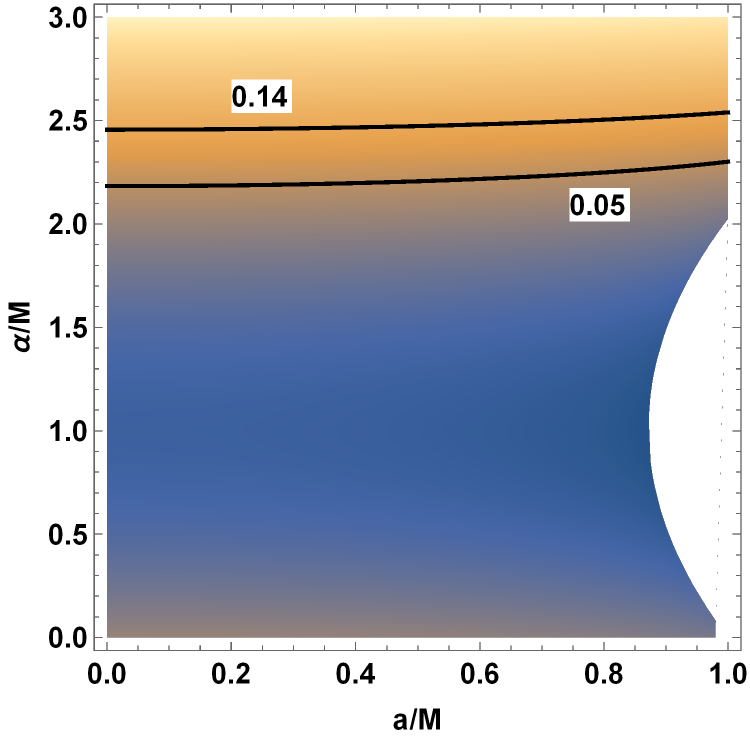}&
\raisebox{.055\height}{\includegraphics[width=0.04\columnwidth]{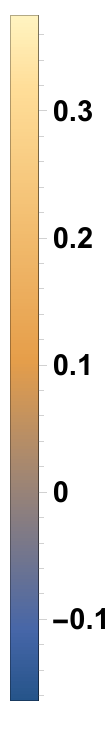}}&
\includegraphics[width=0.35\columnwidth]{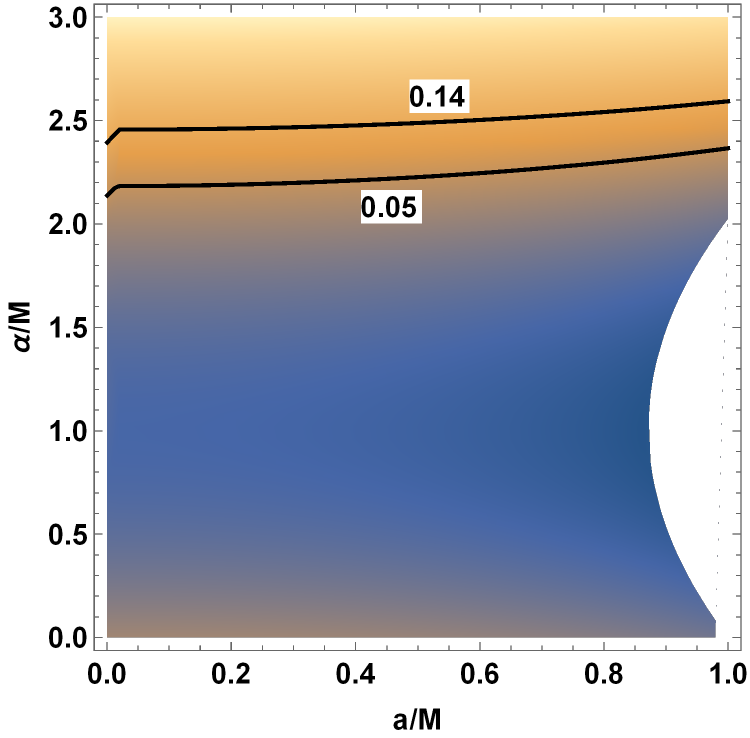}&
\raisebox{.055\height}{\includegraphics[width=0.04\columnwidth]{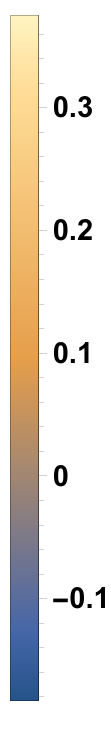}}
\end{tabular}
\caption{Variation of deviation parameter $\d$ with spin $a$ and GUP parameter $\a$. The left one is for $\th_0=90^o$, and the right one is for $\th_o=17^o$. Here we have taken $SgrA^*$ and the bold lines provide bounds given by Keck.}\label{delpara2}
\end{center}
\end{figure}
\section{concluding remark}
GUP provides a novel way to incorporate quantum gravity into GR. Here in this article, we have considered LQG, which produces a minimum measurable length and a maximum measurable momentum. We have carried out our analysis throughout this article with the sole purpose of gauging the impact of GUP on the optical properties of an LQG modified \k BH. A critical value of spin $a=7M/8$ was found owing to the presence of GUP. Below this spin, no extremal BH exists for any real value of GUP parameter $\a$, even though BH solutions exist. This implies we can have an LQKBH for any real value of $\a$ for $a< 7M/8$. However, this changes dramatically as we increase the spin beyond the critical value. Here we have no BH solution for $\a \in (M-\sqrt{8 a M-7 M^2},\, M+\sqrt{8 a M-7 M^2})$. These values of $\a$ are forbidden with regard to the existence of LQKBHs. We get extremal BHs at $\a = M-\sqrt{8 a M-7 M^2}$ and $\a=M+\sqrt{8 a M-7 M^2}$ where the two horizons coincide at the same value for both $\a$ values.\\
We have further probed the impact of $\a$ on the two horizons. For $a< 7M/8$, the event horizon decreases, and the Cauchy horizon increases with $\a$ reaching an extremum value at $\a=1.0M$, after which the trend reverses. However, no such turning point could be found for $a> 7M/8$. In this case, the event horizon decreases and the Cauchy horizon increases with $\a$ for $\a \leq M-\sqrt{8 a M-7 M^2}$ and the reverse is the trend for $\a \geq M+\sqrt{8 a M-7 M^2}$. We then moved on to study the impact of GUP on the shadow of an LQKBH. Here, too, we see a critical point in the shape and size of BH when $a<7M/8$. The size of the shadow initially decreases, and the deviation of shadow from circular shape increases with $\a$ reaching an extremum at $\a=1.0M$. The trend gets reversed beyond $\a=1.0M$. The size (deviation) of the shadow decreases (increases) with $\a$ for $\a \leq M-\sqrt{8 a M-7 M^2}$ and the reverse is the trend for $\a \geq M+\sqrt{8 a M-7 M^2}$. It is evident from Fig. (\r{shadow90}) and (\r{shadow17}) that an LQKBH has a shadow both larger and smaller than that of a \k BH depending on the value of $\a$. A comparison of Fig. (\r{shadow90}) and (\r{shadow17}) further reveals that the deviation of shadow from the circular shape for the inclination angle $17^o$ is less pronounced than that for $90^o$.\\
We finally moved on to this manuscript's central aim: testing our model's viability against the observables related to the shape and size of shadows of SMBHs $M87^*$ and $SgrA^*$ reported by the EHT. These observations provide an unparalleled and unique opportunity to test modified theories of gravity. We have employed bounds on the shadow angular diameter, the deviation from \s radius, and the axial ratio to put constrain on spin and GUP parameters. Even though \k BH agrees well with the observed values, significant uncertainties in the measured bounds leave room for theories of quantum gravity. We have reported upper bounds on $\a$ obtained from observables at $17^o$ and $90^o$ within $1\sigma$ and $2\sigma$ bounds. Within $1\sigma$ interval, upper bounds lie in the range $[2.03921M,\, 2.509971M]$ at $17^o$ and in $[2.03925M,\,2.509975M]$ at $90^o$. Within $2\sigma$ interval, the range is $[2.34149M,\,2.90787M]$ at $17^o$ and in $[2.34164M,\,2.90788M]$ at $90^o$. We can obtain upper bounds on the dimensionless GUP parameter $\a_0$ by considering the Planck mass. This yields following intervals for upper bounds on $\a_0$: $8.01374\times 10^{44}\leq \a_0 \leq 1.491\times 10^{48}$ within $1\sigma$ interval and $9.20205\times 10^{44} \leq \a_0 \leq 1.72737\times 10^{48}$ within $2\sigma$ bounds. The upper bounds reported here are more stringent than those reported in \c{neves, kimet}. Our analysis in this article indicates the viability of LQKBHs as candidates for SMBHs. We are hopeful that with better and more accurate results from the next-generation EHT, we will be able to put more stringent bounds on $\a$. To complement our probe in this article, we will put our model to the test against observed quasiperiodic oscillations for microquasars, which will further enrich our understanding of various facets of quantum gravity.

\end{document}